\documentclass{elsart}

\usepackage{graphics}

\usepackage{amssymb}

\begin{document}

\begin{frontmatter}



\title{Quarkonium at finite temperature}


\author{Takashi Umeda}

\address{Brookhaven National Laboratory, Upton, NY, USA, 11973}

\begin{abstract}
Lattice QCD studies on charmonium at finite temperature are presented
After a discussion about problems for the Maximum Entropy Method 
applied to finite temperature lattice QCD, I show
several results on charmonium spectral functions.
The ``wave function'' of charmonium is also discussed
to study the spatial correlation between quark and anti-quark in
deconfinement phase.

\end{abstract}

\begin{keyword}
Lattice QCD \sep Charmonium \sep QGP
\PACS 12.38.Gc \sep 12.38.Mh
\end{keyword}
\end{frontmatter}

\section{Introduction}
\label{intro}
To investigate the properties of quark gluon plasma (QGP)
in heavy ion collision experiments, theoretical prospects
play important roles, since such processes include complicated
interactions among large number of particles.
Changes of charmonium states have been regarded as one of the most
important probes of plasma formation \cite{NA50},
because the potential model calculations predict the mass shift
of charmonium near $T_c$ \cite{Has86},
and $J/\psi$ suppression above $T_c$ \cite{Mat86}.
However, lattice QCD simulations have indicated that the
thermal properties of hadronic correlators are much more
involved than weakly interacting almost free quarks \cite{TARO01}.
Furthermore, recent lattice QCD studies of spectral functions of
charmonium suggest that hadronic excitations of $c$-$\bar{c}$ system
may survive even above $T_c$ \cite{Ume03,Pet03,Asa03}.

In this proceeding Lattice QCD studies for properties of charmonia
at finite temperature are presented. 
In the Sect.\ref{spfunc} the charmonium spectral function
are discussed. There are several difficulties to extract the spectral
function from temporal correlation function at finite temperature.
After I summarize the difficulties and explain how to overcome it,
several numerical results are presented.
In the Sect.\ref{wavefunc} I show a study of ``wave function'' of
charmonium at finite temperature. 
Although the study was performed before a serious
of the studies of spectral function, it already shows clear
nonperturebative nature charmonium state in the QGP.
Finally I summarize this proceeding.
\section{Spectral function of charmonium at finite temperature}
\label{spfunc}
\subsection{Correlation function of charmonium at finite temperature.}
For the study of charmonium physics at $T>0$ on a lattice, there are
several problems. Then I classify these problems into two
categories.
\begin{description}
 \item[(i)] Precise calculation of temporal correlator of charmonium at
	    $T>0$ 
 \item[(ii)] Extraction of physical properties of charmonium from the 
	    correlators 
\end{description}
First I consider the former one, (i).
In the lattice QCD simulation at $T>0$,
we set a temporal lattice extent to $1/T$.
At high temperature, one needs the large lattice cutoff
to work with the sufficient degrees of freedom in the 
temporal direction.
In order to obtain the detailed information of temporal 
meson correlators at $T>0$, a high resolution in temporal
direction is needed.
The large lattice cutoff is also necessary to study a correlator
of meson with the heavy quarks because of $O(m_q a)$ error in quark
action.
If one tries to overcome these difficulties with straightforward way,
the tremendous large computational power is necessary.
In order to get the sufficiently fine resolution
with limited computer resources, the anisotropic lattice,
which has a finer temporal lattice spacing $a_{\tau}$ than
the spatial one $a_{\sigma}$, is a reasonable solution.

\begin{figure}
\begin{center}
\resizebox{140mm}{!}{
 \includegraphics{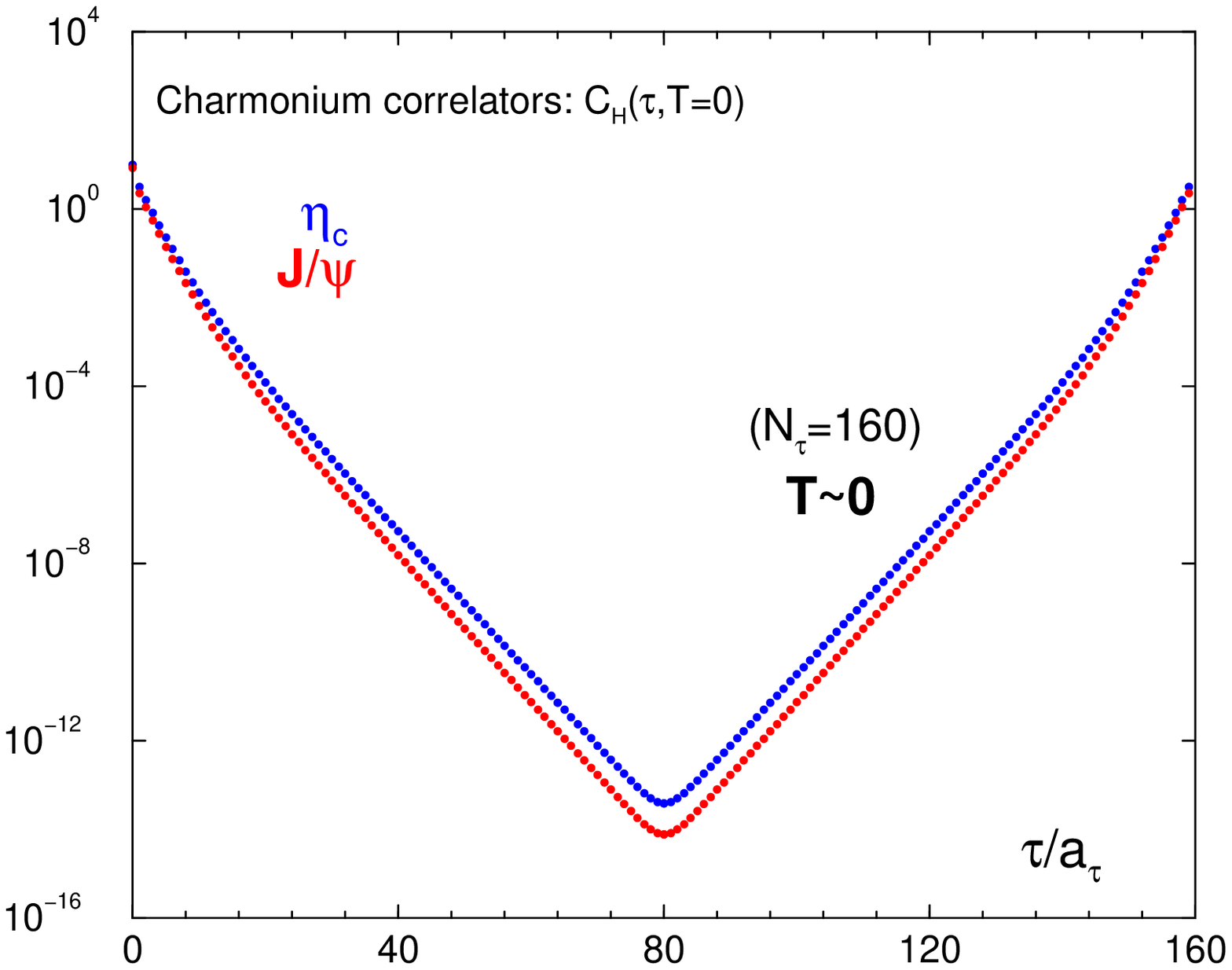}
 \includegraphics{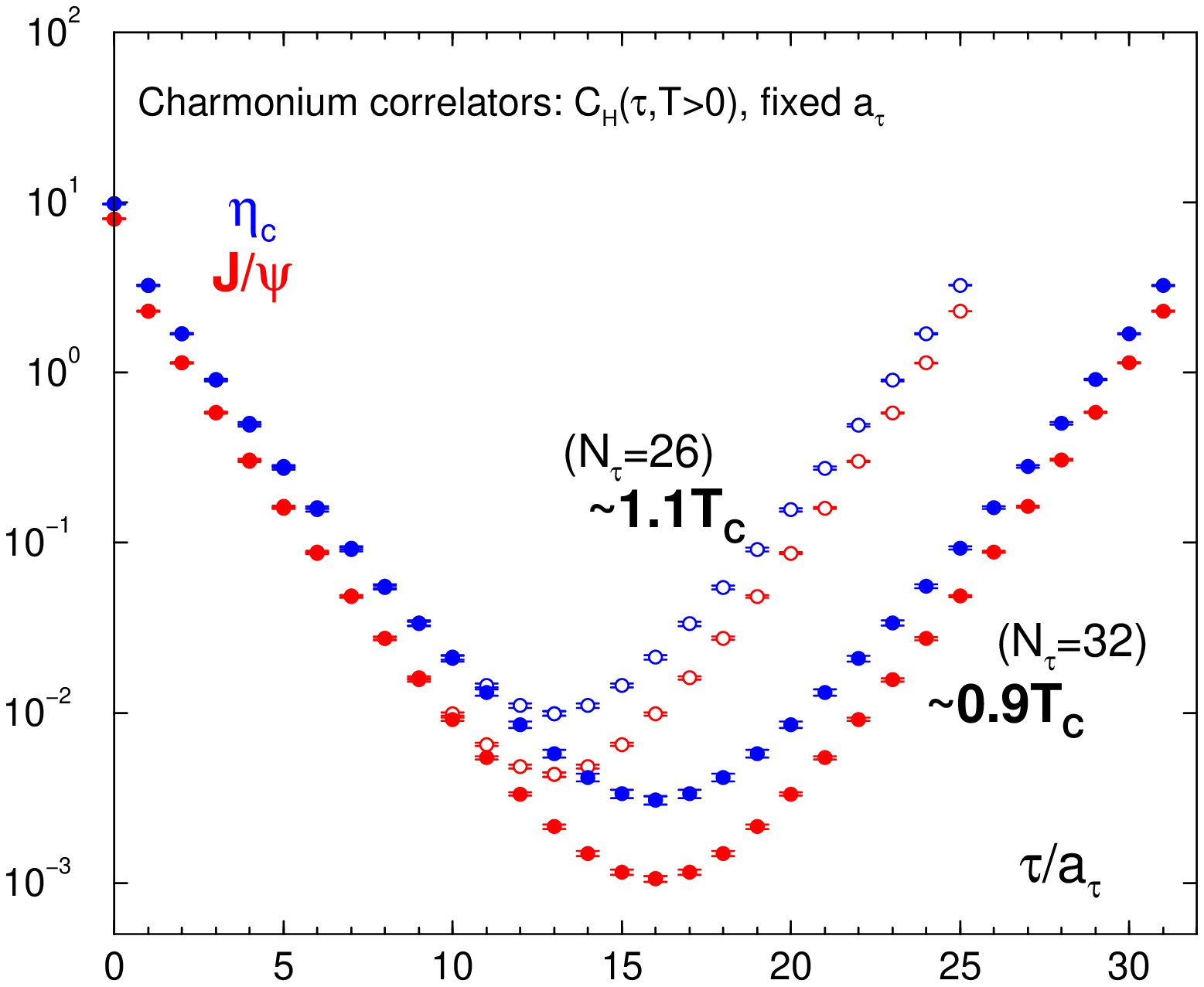}
}
\caption{Samples of charmonium correlation function at $T=0$ (left) and
 at $T>0$ (right). Vertical axis is logarithmic scale. Details of
 numerical setup is described in Sect.\ref{sec:numerical}} 
\label{fig_corr}
\end{center}
\end{figure}

Here I show, in fig.\ref{fig_corr}, samples of charmonium correlation
function using the anisotropic lattice, whose numerical setup is
described in Sect.\ref{sec:numerical}.
In this case, there are 160 data points in temporal direction at $T=0$
On the other hand, at $T>0$ near the $T_c$
the number is reduced to about 30 data points
or less, furthermore only half data points are available 
because of the (anti)periodic boundary condition in t-direction.
Due to restriction of computational resources, it is difficult 
to largely increase the data point than that of the sample even in
quenched simulation. 

The latter problems, (ii), are much more difficult to be overcome.
This is because one is inevitably enforced to extract the spectral
properties from the data at the short temporal distance, where high
frequency component of the dynamics is significant.
In order to extract the information of spectral function from finite
temperature correlation function in such a condition the Maximum Entropy
Method (MEM) is proposed.
The method successfully works in zero temperature lattice QCD
\cite{nakahara99,yamazaki}. 
In the next section I discuss the method applied to finite
temperature system. 

\subsection{Maximum entropy method}
Let me briefly summarize the outline of MEM basically following
Ref.~\cite{nakahara99}, which reviews in detail MEM applied to data
of lattice QCD simulation.
One obtain the spectral function, $A(\omega)$, from the 
given lattice result for the correlator, $C(t)$, 
by solving the inverse problem, 
\begin{equation}
C(t) = \int_0^{\infty} d\omega K(t,\omega) A(\omega),
\end{equation}
where the (continuum type) kernel $K(t,\omega)$ is given by
\begin{equation}
 K(t,\omega)=\frac{ e^{-\omega t}+e^{-\omega(N_t-t)} }
                     { 1 - e^{-N_t\omega} }.
\label{eq:kernel}
\end{equation}
To extract the $A(\omega)$, MEM maximizes a
functional $Q(A;\alpha) = \alpha S[A] - L[A]$.
$L[A]$ is the usual likelihood function,
and minimized in the standard $\chi^2$ fit.
The Shannon-Jaynes entropy $S[A]$ is defined as
\begin{equation}
 S[A] = \int_0^\infty d\omega
   \left[ A(\omega)-m(\omega)-A(\omega)
           \log \left( \frac{A(\omega)}{m(\omega)}\right) \right].
\end{equation}
The function $m(\omega)$ is called the default model function, 
and should be given as a plausible form of $A(\omega)$.
At the last stage of calculation the parameter $\alpha$ can be
integrated out by a weighted average of prior probability for $\alpha$. 

In the maximization step of $Q(A;\alpha)$ the singular value
decomposition of the kernel $K(t,\omega)$ is usually used
\cite{nakahara99}.\footnote{
Analysis of MEM without singular value decomposition was examined
in Ref.~\cite{Fiebig:2002sp}.}
Then the SPF is represented as a linear
combination of the eigenfunctions of $K(t,\omega)$:
\begin{eqnarray}
 A(\omega)=m(\omega)
\exp\left\{ \sum_{i=1}^{N_s} b_i u_i(\omega) \right\},
\label{eq:spec_svd}
\end{eqnarray}
where $N_s$ is the number of eigenfunctions, $b_i$ are parameters, and 
$u_i(\omega)$ the eigenfunction of the kernel $K(t,\omega)$.
The number of degrees of freedom of $A(\omega)$ is accordingly
reduced to the number of data points of the correlator.
Although $b_i$
can in principle be determined uniquely from the data without
introducing an entropy term, the small eigenvalues of $K(t,\omega)$
lead to a singular behavior of the spectral function;
hence truncation of the terms is practically required,
i.e. $N_s$ may be less than the number of data points
\cite{taro98}.  
In MEM, the entropy term stabilizes the problem and
guarantees an unique solution for the coefficients of the eigenfunctions
\cite{nakahara99}.

An outstanding feature of Eq.(\ref{eq:spec_svd}) is that
it can be fitted to generic shape without restriction to
specific forms such as a sum of poles.
However, the resolution of course depends on the number of degree of
freedom in Eq.~(\ref{eq:spec_svd}), and also on $\omega$.

As mentioned in this section, 
MEM needs a default model function to define the entropy term $Q$.
Since small difference between a trial SPF and default
model function makes the entropy term large, 
the default model function strongly affects the result of MEM
when the quality of data is not sufficient.
Therefore the default model function should include only reliable
information we know beforehand. 
If not so, there is a risk the result might be controlled by hand.

\subsection{Numerical results of the charmonium spectral functions}
\label{sec:numerical}
Although the MEM applied to finite temperature has some subtleties,
it is one of the most powerful tools to study spectral functions.
Therefore there are many studies of the charmonium spectral function by
using the MEM. 
Here I present some results of the studies, based on our works.

The first studies is performed on quenched anisotropic lattice.
To avoid the subtlety of the MEM the authors adopt the MEM and
the standard least square analysis with an ansatz for shape of spectral
functions.
The result of the former provides the prior knowledge required
for the latter.
Combining the two methods, the results may be more
reliable and quantitative than with one of them.
In the previous study \cite{Ume03} suggested that MEM fails to
extract the spectral function from the correlator of local
operators at $T>0$ in their lattice setup.
Therefore spatially extended operators are adopted to 
enhance the low frequency mode of the spectral function.
However the smeared operators may lead to an artificial peak,
and thus careful analysis is necessary to distinguish the
physical results from the artifact ones.

The simulation is performed on lattices with the spatial
lattice cutoff
$a_\sigma^{-1}\simeq 2$ GeV and the anisotropy $a_\sigma/a_\tau=4$
in the quenched approximation \cite{matsufuru01}.
The lattice sizes are $20^3\times N_t$, where $N_t$ are from $160$
($T\simeq 0$) to $16$ ($T\simeq 1.75 T_c$).
The numbers of configurations are 500 at $T\simeq 0$ and 1000 
at $T>0$.
$N_t=28$ roughly corresponds to the transition temperature.
The quark field is described by the $O(a)$ improved
Wilson action with the tree-level tadpole-improvement.
The hopping parameter is chosen so that the charmonium spectrum
is roughly reproduced.

\begin{figure}
\begin{center}
\resizebox{140mm}{!}{
 \includegraphics{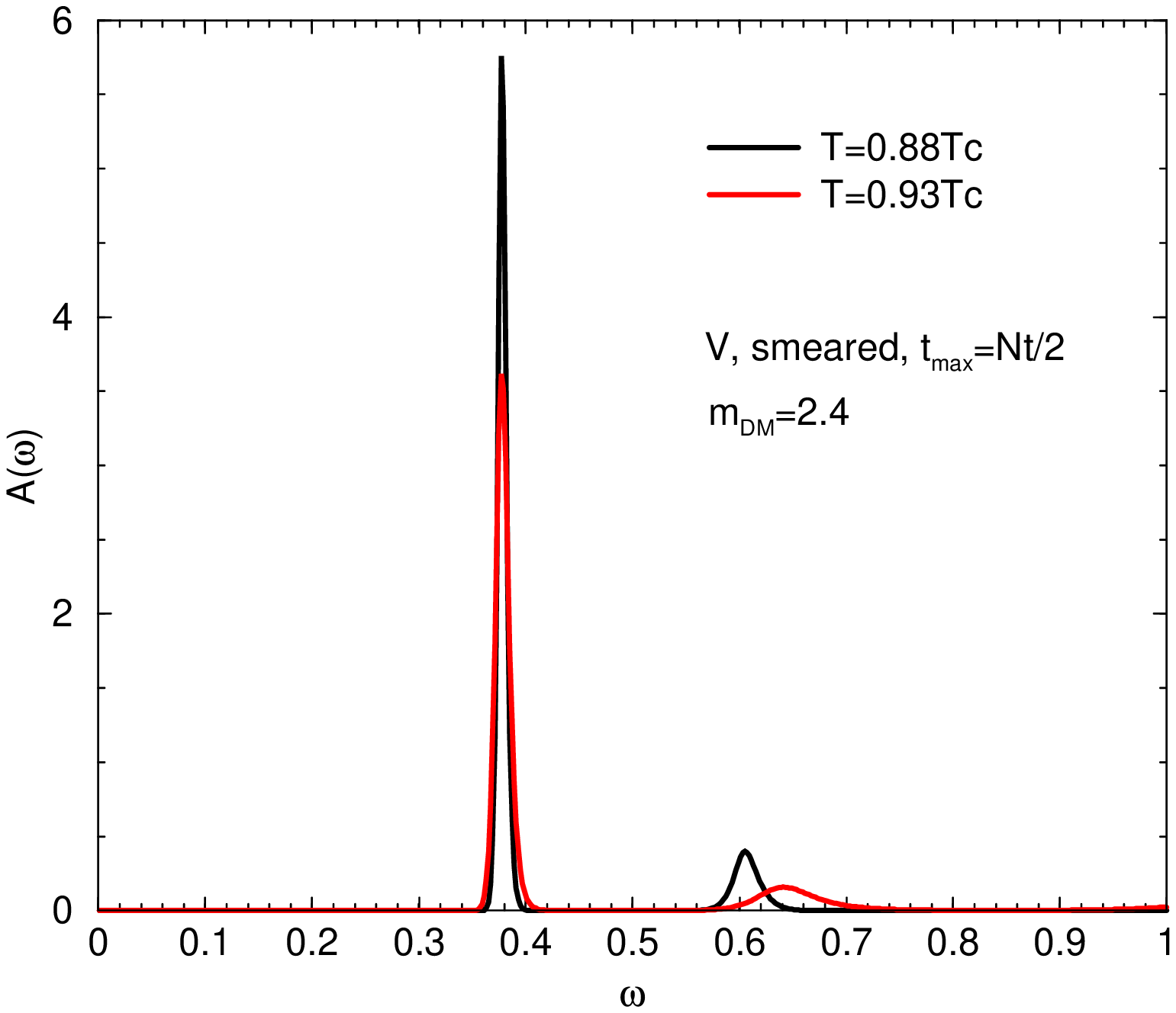}
 \includegraphics{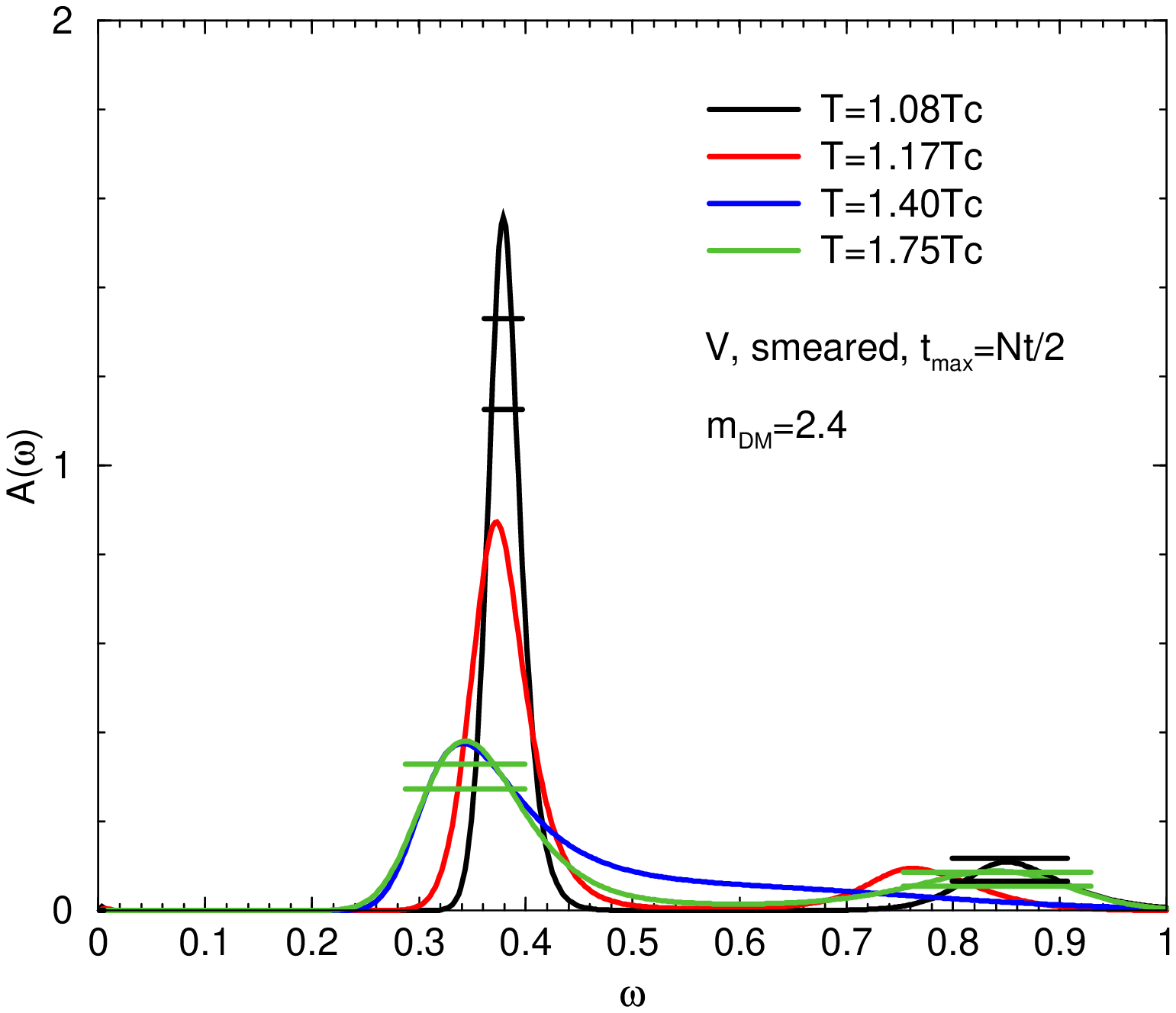}
}
\caption{Charmonium spectral function from a smeared operators in V
 channel, whose lowest state corresponds to $J/\psi$, below $T_c$ (left)
 and above $T_c$ (right) \cite{Ume03}. Lines around peaks
 present an error of the peak in the MEM.}
\label{fig:ume}
\end{center}
\end{figure}

Although the MEM result shows that 
the spectral functions have peak structure in PS and V channels
(corresponding to $\eta_c$ and $J/\psi$) at all temperature, 
the difference of result for different smearing functions,
``smeared'' and ``half-smeared'', exists at higher temperature,
especially above $1.4T_c$.
This means that the peak structure of the spectral function at high
temperature might be artificial.
Furthermore they find large default model function dependence of the
results, in which the position of the peak is stable but the peak width
has large dependence.
Therefore they conclude that the peak structure of the spectral function
at higher temperature, i.e. more than $1.4T_c$, may be artificial.
However, up to $1.4T_c$ the peak structure, in other words, hadronic
excitations may survive even in the deconfinement phase.

\begin{figure}
\begin{center}
\resizebox{140mm}{!}{
 \includegraphics{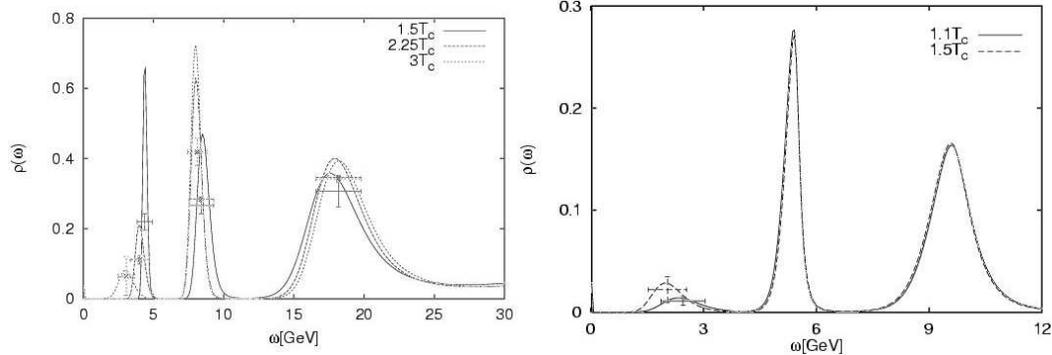}
}
\caption{Charmonium spectral function above $T_c$ in V channel, whose
 lowest state corresponds to $J/\psi$ (left), and in AV channel, whose
 lowest state corresponds to $\chi_c$ (right) \cite{Pet03}.}
\label{fig:bielefeld}
\end{center}
\end{figure}

Next the Bielefeld group tried to calculate on isotropic
quenched lattices with local operators \cite{Pet03}.
Figure \label{fig:bielefeld} (left) shows the spectral function in V
channel at various temperature above $T_c$.
Their result also shows that there are peak structures in PS and V
channels above $T_c$. The peaks survive even above $2T_c$.
Furthermore they calculated the axial vector channel, fig.\label{fig}
(right), whose lowest peak corresponds to $\chi_c$ state. 
The correlation function of the axial vector channel shows drastic
change just above $T_c$, then the lowest peak of the spectral function 
disappears just after the deconfinement phase transition.
Hatsuda and Asakawa studied on anisotropic quenched lattice with local
operators, and their results also show the existence of the hadronic
state up to about $1.6T_c$ \cite{Asa03}.
The details of the result were presented in this conference.
Recently the Trin-lat group calculated on anisotropic 2-flavor full QCD
lattice. The result also consistent with all previous quenched lattice
results. 

\section{Wave function of charmonium at finite temperature}
\label{wavefunc}

In this subsection the $c\bar{c}$ bound state at $T>0$ is discussed in
the light of ``wave function''.  
The definition of the ``wave function'' in the Coulomb gauge is as
follows. 
\begin{equation}
 w_M (\vec{r},t) = \sum_{\vec{x}} \langle
   \bar{q}(\vec{x}+\vec{r},t)\gamma_M
   q(\vec{x},t)O_M^\dagger(0)\rangle
\end{equation}
Here this definition is the same form as the Bethe-Salpiter function.  
This wave function shows the spatial correlation between $q$ and
$\bar{q}$, and gives us a hint of the mesonic bound state from its
$t$ dependence. 
In the case of free quarks, $q \bar{q}$ has no bound state, then the
wave function ought to broaden with $t$. 
On the other hand if 
quark and anti-quark form a bound state, 
the wave function holds a stable shape with $t$.
We can discuss the existence of such a bound state by observing the
$t$-dependence of the the wave function.
For this purpose we compare the correlation at spatial
origin with another spatially separated point at each $t$.
Therefore we define the wave function normalized at the spatial origin,
$\phi_M(\vec{r},t)$, as follows,
\begin{equation}
\phi_M(\vec{r},t) =  
  \frac{w_M(\vec{r},t)}{w_M(\vec{0},t)}.
\end{equation}
From now on the wave function denotes this normalized definition.
\begin{figure}
\begin{center}
\resizebox{80mm}{!}{
 \includegraphics{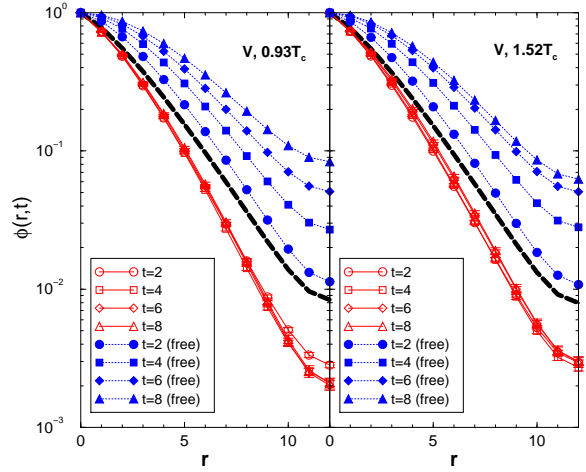}
}
\caption{The spatial quark and antiquark correlation below and above
 $T_c$ \cite{Ume00}. The dashed lines are an initial distribution of
 quark and antiquark. The solid symbols are result in free quark case.}
\label{fig:wavefunc}
\end{center}
\end{figure}

Fig.~\ref{fig:wavefunc} shows the results at $T>0$
with the smeared source function which is slightly wider than the observed
wave function at $T=0$. 
The wave functions composed of free quark propagators are also shown
together.  

As is shown in the Fig.~\ref{fig:wavefunc}, the behaviors of the
observed wave functions are clearly different 
from that of the free quark case at each temperature and in each mesonic  
channel. In the free quark case, as I expected, the wave functions are
broadening as $t$ increased. 
On the other hand, the observed wave functions hold stable shape which
is the slightly narrower than source function.
These behaviors are independent of the source function.

\section{Summary}
\label{summary}

In this proceeding I present the lattice QCD studies of charmonia at
finite temperature. 
The studies of spectral functions of charmonium at finite temperature
is possible by using the MEM. 
The MEM is powerful tool to study the spectral function, but various
systematic uncertainties have to be controlled.
Most study of the spectral function is consistent with 
an existence of $\eta_c$ and $J/\psi$ up to 1.5-2 $T_c$.
Other approach such as ``wave function'' also consistent with a 
compact state composed of charm and anti-charm quarks even above $T_c$.
Although these conclusion is rather strange in view of a naive QGP
picture, our lattice QCD results and other phenomenological studies
support this interesting strongly interacting QGP picture.

As next steps with lattice QCD simulations, 
accurate determination of the dissociation temperature is required.
For the purpose reliable calculations with full QCD simulation are
necessary. 
For phenomenological discussion of $J/\psi$ suppression,
$\psi'$ and $\chi_c$ states should be studied in detail.


\begin{thebibliography}{00}

\bibitem{NA50} NA50 Collaboration,
   Phys. Lett. B 477 (2000) 28.

\bibitem{Has86} T.~Hashimoto et al,
  Phys. Rev. Lett. 57 (1986) 2123.

\bibitem{Mat86} T.~Matsui and H.~Satz,
  Phys. Lett. B 178 (1986) 416.

\bibitem{TARO01} QCD-TARO Collaboration, 
  Phys. Rev. D {\bf 63} (2001) 054501;

\bibitem{Ume03} T.~Umeda, {\it et al.},
  Eur.\ Phys.\ J.\ C {\bf 39S1}, 9 (2005)
  [arXiv:hep-lat/0211003].

\bibitem{Pet03} S.~Datta et al.,
   J. Phys. G30 (2004) S1347.

\bibitem{Asa03} M.~Asakawa and T.~Hatsuda,
   Phys. Rev. Lett. 92 (2004) 012001.

\bibitem{nakahara99} M.~Asakawa {\it et al.},
Prog.\ Part.\ Nucl.\ Phys.\  {\bf 46}, 459 (2001).

\bibitem{yamazaki} T.~Yamazaki {\it et al.}  [CP-PACS Collaboration],
Phys.\ Rev.\ D {\bf 65}, 014501 (2002).

\bibitem{Fiebig:2002sp} H.~R.~Fiebig,
Phys.\ Rev.\ D {\bf 65} (2002) 094512.

\bibitem{taro98} For pioneering works,
 QCD-TARO Collaboration, Ph. de Forcrand {\it et al.},
  Nucl.\ Phys.\ B (Proc. Suppl.) {\bf 63}, 460 (1998);
 E.~G.~Klepfish, C.~E.~Creffield, E.~R.~Pike,
  {\it ibid.}, 655 (1998).

\bibitem{matsufuru01} H.~Matsufuru {\it et al.}, 
Phys.\ Rev.\ D {\bf 64}, 114503 (2001).

\bibitem{Ume00} T.~Umeda {\it et al.},
  Int. J. Mod. Phys. A {\bf 16} (2001) 2215.

\end{thebibliography}
\end{document}